\documentclass[preprint,10pt,5p,times,twocolumn]{elsarticle}
\usepackage{amsfonts}
\usepackage{amsmath} % math
\usepackage{amssymb} % math
\usepackage{graphicx}
\usepackage{color}
\journal{Physics Letters A}
\begin{document}
\title{Amplitude modulated drift wave packets in a nonuniform magnetoplasma}
%\author{P. K. Shukla}
%\email{profshukla@yahoo.de}
%\affiliation{International Centre for Advanced Studies in Physical Sciences \& Institute for Theoretical Physics,
%Faculty of Physics and Astronomy, Ruhr University Bochum, D-447 80 Bochum, Germany}
%\affiliation{Department of Mechanical and Aerospace Engineering \& Center for Energy Research,
% University of California San Diego, La Jolla, CA 92093, U. S. A.}
%\author{A. P. Misra}
%\email{apmisra@gmail.com}
%\affiliation{Department of Mathematics, Visva-Bharati University, Santiniketan-731 235, India}
\author[PKS]{P. K. Shukla\fnref{fn1}}
\ead{profshukla@yahoo.de}
\author[APM]{A. P. Misra\corref{cor1}}
\ead{apmisra@gmail.com}
\cortext[cor1]{Corresponding author.}
\fntext[fn1]{ Also at Department of Mechanical and Aerospace Engineering \& Center for Energy Research, University of California San Diego, La Jolla, CA 92093, U. S. A. }
 \address[PKS]{International Centre for Advanced Studies in Physical Sciences \& Institute for Theoretical Physics, Faculty of Physics and Astronomy, Ruhr University Bochum, D-447 80 Bochum, Germany}
\address[APM]{Department of Mathematics, Siksha Bhavana, Visva-Bharati University, Santiniketan-731 235, India}
 %\received{18 June 2012}
%\date{18 June 2012}
\begin{abstract}
We consider the amplitude modulation of  low-frequency, long wavelength electrostatic drift wave packets in 
a nonuniform magnetoplasma with the effects of equilibrium density, electron temperature and magnetic field inhomogeneities. The dynamics of the modulated drift wave packet is governed by a nonlinear Schr\"odinger equation. The latter is used to study the modulational instability of a Stoke's wave train to  a small longitudinal perturbation. It is shown that  the drift wave packet is stable (unstable)  against the  modulation   when the drift wave number lies in $0< k < 1/\sqrt{2}$  $(1/\sqrt{2}<k<1)$. Thus, the modulated drift wave packet can propagate in the form of bright and dark envelope solitons or as a drift wave rogon.
\end{abstract}
%\pacs{52.27.Aj, 50.35.Kt, 50.35.Sb, 52.35.Mw}
\maketitle
  
A nonuniform magnetoplasma supports a great variety of low-frequency electrostatic and electromagnetic drift-type 
modes. Examples include the electrostatic drift waves \cite{r1} and coupled drift-Alfv{\'e}n waves \cite{r2},
which play a crucial role in cross-field plasma particle transports \cite{r3,r4}, and the formation of coherent structures 
in space \cite{r5} and laboratory \cite{r6,r6a} plasmas that are magnetized. Both the
drift-Alfv\'en waves can be excited by free energy sources that are stored in the equilibrium pressure gradient
and in magnetic field inhomogeneity. Nonthermal drift waves attain large amplitudes and start interacting among themselves.
In the past, Tasso \cite{r7} and Orevskii {\it et al.} \cite{r8} considered nonlinear interactions between one-dimensonal 
drift waves propagating in a direction orthogonal to the density and temperature gradients and a uniform magnetic 
field in an electron-ion plasma. They reported the formation of non-envelope drift solitary pulses that were used 
in the study of drift wave turbulence comprising an ensemble of drift wave solitons \cite{r3,r9} in magnetized plasmas. 

Hasegawa and Mima \cite{r10} incorporated the vector nonlinearity associated with the nonlinear ion polarization drift in 
the study of nonlinearly interacting pseudo-three-dimensional drift waves in a nonuniform plasma without the electron
temperature and magnetic field inhomogeities. The Hasegawa-Mima equation, which similar to the Charney equation 
\cite{r10a} governing the dynamics of the Rossby waves in the atmosphere, admits a Larichev-Rezhnik type-dipolar 
vortex \cite{r3,r10b,r11,r12}  as one of the possible stationary states of the drift wave turbulence. The mode 
couplings between finite amplitude drift waves can  also generate convective cells and zonal flows \cite{r12a}. 
The latter provide a better plasma confinement, since they act as a barrier  for inhibiting the transport of the plasma 
particles across the external magnetic field direction.  Tynan {\it et al.} \cite{r12b} have presented an elegant review of experimental  drift wave turbulence studies.

In this Letter, we discuss the properties of modulated one-dimensional drift wave packets in a nonuniform magnetoplasma 
with the effects of equilibrium density, electron temperature and magnetic field gradients. It is shown that the dynamics of 
the modulated drift wave packet is governed by a nonlinear Schr\"odinger equation (NLSE), which depicts the formation of dark 
and bright solitons, as well as drift rogue waves (or drift rogons).         

Let us consider a nonuniform electron-ion plasma in the presence of equilibrium density and electron temperature 
inhomogeneities in a nonuniform external magnetic field $\hat {\bf z} B_0 (x)$, where $\hat {\bf z}$ is the 
unit vector along the $z-$ axis of a Cartesian coordinate system, and $B_0$ is the strength of the magnetic field. Thus, at equilibrium, we have

%1%eqbm
\begin{equation}
\frac{\partial}{\partial x}\left[n_0 (x) T_e (x) + \frac{B_0^2(x)}{8\pi}\right]=0,
\label{eqbm}
\end{equation}
where $n_0 (x)$ and $T_e (x)$ are the unperturbed electron number density and the electron temperature,
which have gradients along the $x$-axis.

In the presence of the low-frequency (in comparison with the ion gyrofrequency $\omega_{ci} =eB_0/m_i c$,
where $e$ is the magnitude of the electron charge, $m_i$ is the ion mass, and $c$ is the speed of light in vacuum),
long wavelength (in comparison with the ion thermal gyroradius $\rho_i =V_{Ti}/\omega_{ci}$, where 
$V_{Ti} =(k_B T_i/m_i)^{1/2}$ is the ion thermal speed, $k_B$ is the Boltzmann constant, and $T_i$ is the ion temperature) 
electrostatic field $({\bf E}=-\nabla \phi)$, where $\phi$ is the electrostatic potential  of the drift waves, the perpendicular 
(to $\hat {\bf z}$) component of the electron and ion fluid velocities are,  respectively,

%2 ev 
\begin{eqnarray}
&&{\bf u}_{e\perp} = \frac{c}{B_0(x)}\hat {\bf z} \times \nabla \phi -\frac{c k_B T_e (x)} 
{eB_0(x)n_0 (x)}\hat {\bf z} \times \nabla n_{e1} \notag \\
&&\hskip20pt \equiv {\bf V}_E + {\bf V}_{De}, \label{ev}\\
%3 iv
&&{\bf u}_{i\perp} = \frac{c}{B_0(x)}\hat {\bf z} \times \nabla \phi-\frac{c}{B_0 (x) \omega_{ci}}\left(\frac{\partial}{\partial t} 
+{\bf V}_E \cdot \nabla\right) \nabla_\perp \phi \notag\\
&& \hskip20pt\equiv {\bf V}_E + {\bf V}_p, \label{iv}
\end{eqnarray}
where we have assumed that $|d/dt| \ll \nu_{ei} \ll \omega_{ce}$, with $ d/dt \equiv \partial/\partial t + {\bf V}_B \cdot \nabla$.
Here $\nu_{ei}$ is the electron-ion collision frequency, $ \omega_{ce} =eB_0/m_e c$ is the electron gyrofrequency, and $m_e$ is the electron mass. Furthermore, without loss of generality, we have taken $T_i \ll T_e$. The electron density
perturbation is denoted by $n_{e1}$ $(\ll n_0)$.        
   
Inserting Eq. \eqref{ev} into the electron continuity equation and using the parallel component of the inertialess electron 
momentum equation with $|d n_{(e1)}/dt| \ll n_{ei} n_{e1}$, we obtain for $\nu_{ei} |d n_{e1}/dt| \ll V_{Te}^2 |\partial^2 n_{e1}/\partial z^2|$
and $c [\partial (n_0/B_0)/\partial x] \partial^2 \phi/\partial y \partial t \ll (n_0 e k_B T_e /\nu_{ei})| \partial^2 \phi/\partial z^2|$,
the Boltzmann law for the electron number density perturbation \cite{r11}
%4 eB
\begin{eqnarray}
&&n_{e1} = n_0 (x) \exp \left[\frac{e\phi}{k_B T_e(x)}\right] \notag\\
&&\hskip20pt \approx\frac{n_0 (x)}{k_B T_e (x)} e \phi + \frac{1}{2} \left( \frac{n_0 (x)}{k_B T_e (x)} \right)^2 e^2 \phi^2 
\label{eB}.
\end{eqnarray}  
Furthermore, substituting Eq. \eqref{iv} into the ion continuity equation, we obtain
%5 ion5
\begin{equation}
\frac{\partial n_{i1}}{\partial t} + \nabla \cdot [(n_0 (x) + n_{e1}) {\bf V}_E ] 
\approx \frac{c n_0}{B_0 \omega_{ci}} \frac{d \nabla_\perp^2 \phi}{dt}
\label{ion5},  
\end{equation}
where we have neglected the parallel (to $\hat {\bf z})$ ion dynamics, thereby discarded the coupling 
between the drift and ion-sound waves. 

We can  now combine Eqs. \eqref{eB} and \eqref{ion5}  under the quasi-neutrality condition $n_{i1} = n_{e1}$, which holds for a magnetized plasma
with $\omega_{pi} \gg \omega_{ci}$, where $\omega_{pi} =(4\pi n_0e^2/m_i)^{1/2}$ is the ion plasma frequency, to obtain the 
drift wave equation in one-space dimension 

%6 driftwave
\begin{eqnarray}
&& \left(1-\frac{\partial^2}{\partial y^2}\right)\frac{\partial \varphi}{\partial t}+\alpha \frac{\partial \varphi}{\partial y}
+ \varphi \frac{\partial \varphi}{\partial t} -\gamma \varphi \frac{\partial \varphi}{\partial y}=0, 
\label{driftwave}
\end{eqnarray} 
where $\varphi=e\phi/k_B T_e$, $\alpha =\rho_s/L_n$, and $\gamma= \rho_s/L_{TB}$. Here $\rho_s =C_s/\omega_{ci}$ is the ion 
sound gyroradius, $C_s =(k_B T_e/m_i)^{1/2}$ is the ion sound speed, $L_n^{-1} = - \partial {\rm ln} (n_0/B_0)/\partial x >0$, and 
$L_{TB} = \partial {\rm ln }(n_0/T_e B_0)/\partial x$. Furthermore, the time and space variables are in units of the  ion gyroperiod $1/\omega_{ci}$ and $\rho_s$.  

Let us now derive the governing nonlinear equation for amplitude-modulated drift wave packets, following the standard multiple-scale
technique \cite{r13,r14,r15}. Then in a coordinate frame moving with the speed $v_g$, the space and the time variables can be stretched 
as $\xi=\epsilon (y-v_gt)$, $\tau=\epsilon^2 t$, where $\epsilon$ is a small parameter ($0 < \epsilon \ll 1)$ representing the weakness of perturbation. We are interested in the 
modulation of a plane drift wave as the carrier wave with the wave number $k$ and the frequency $\omega$. The dynamical variable 
$\varphi$ can be expanded as

%7 expansion
\begin{equation} 
\varphi=\sum^{\infty}_{n=1}\epsilon^n \sum^{\infty}_{l=-\infty}\varphi_l^{(n)}(\xi,\tau)\exp[i(k y-\omega t)], 
\label{expansion}
\end{equation} 
where $\varphi_{-l}^{(n)}=\varphi_{-l}^{(n)*}$ is the reality condition and the asterisk denotes the complex conjugate.

Substituting the expansion, given by Eq. \eqref{expansion}, and the stretched coordinates in Eq. \eqref{driftwave}, and equating different 
powers of $\epsilon$, we obtain for $n=l=1$ (coefficient of $\epsilon$) the  linear dispersion relation for drift waves 

%8 DR
\begin{equation}
\omega =\frac{\alpha k}{1+k^2}.
\label{DR}
\end{equation}

From the second-order expressions for $n=2,l=1$ we obtain an equation in which the coefficient of $\varphi_1^{(2)}$ vanishes due to the dispersion relation, 
and the coefficient of $\partial \varphi_1^{(1)}/\partial\xi$, after equating to zero, gives the  group velocity 

 %9 vg
\begin{equation} 
%v_g=\frac{\alpha-2\omega k}{1+k^2}=\frac{\alpha(1-k^2)}{(1+k^2)^2}. \label{vg}
v_g=\frac{\alpha(1-k^2)}{(1+k^2)^2}. \label{vg}
\end{equation} 

Next, the zeroth harmonic mode, which appears due to the nonlinear self-interaction of the carrier waves in the coefficient 
of $\epsilon^3$ for $n=2,l=0$, is obtained as

%10 zerothorder
\begin{equation}
\varphi_0^{(2)}=\left(\frac{\gamma+v_g}{\alpha-v_g}\right)|\phi_1^{(1)}|^2. \label{zerothorder}
\end{equation}
Considering the second order harmonic mode for $n=l=2$,  we obtain from the coefficient of $\epsilon^2$

%11 2ndorder
\begin{equation}
\varphi_2^{(2)}=\frac{\omega+\gamma k}{2 \alpha k-\omega(1+4k^2)}[\phi_1^{(1)}]^2.\label{2ndorder}
\end{equation}

 Finally, for $n=3, l=1$, we obtain an equation for third-order first-harmonic modes in which the coefficients of $\varphi^{(3)}_1$ 
and $\partial \varphi_1^{(2)}/\partial\xi$ vanish by the dispersion relation and the group velocity, respectively. 
In the reduced equation we substitute the expressions for $\varphi_0^{(2)}$ and $\varphi_2^{(2)}$ from Eqs. \eqref{zerothorder} and \eqref{2ndorder}. Thus, we obtain the 
following NLSE  
%12 NLSE
\begin{equation}
i\frac{\partial \Phi}{\partial \tau}+P\frac{\partial^2 \Phi}{\partial \xi^2}+Q|\Phi|^2\Phi=0, \label{NLSE}
\end{equation}
where $\Phi=\varphi^{(1)}_1$ is the potential perturbation, or in terms of the original variables

%13
\begin{equation}
i\left(\frac{\partial }{\partial t}+v_g\frac{\partial}{\partial y} \right)\Phi 
+ P\frac{\partial^2 \phi}{\partial y^2}+Q|\Phi|^2\Phi=0,
\end{equation}
where $\Phi\sim \epsilon \varphi^{(1)}_1$. The coefficients of the drift wave group dispersion and the nonlinearity are

%14 P-dispersion
\begin{equation}
%P \equiv\frac{1}{2}\frac{\partial^2 \omega}{\partial k^2}=-\frac{\omega+2kv_g}{1+k^2} 
%= -\frac{\alpha k(3-k^2)}{(1+k^2)^3},\label{P-dispersion}
P \equiv\frac{1}{2}\frac{\partial^2 \omega}{\partial k^2}= -\frac{\alpha k(3-k^2)}{(1+k^2)^3},\label{P-dispersion}
\end{equation}
and
%15
\begin{equation}
Q=\frac{[\alpha+\gamma(1+k^2)]Q_0}{\alpha k(1+k^2)^2(3+k^2)(1-2k^2)}, 
\label{Q-nonlinear}
\end{equation}
where $Q_0=\alpha(1+3k^4)+\gamma\left[(1+k^2)^2+k^2(1-k^4)\right]$.

The propagation of wave packets in a dispersive nonlinear plasma medium   has been known to be subjected to the amplitude modulation, i.e., a slow variation of the wave
packet’s envelope due to the nonlinear self-interaction of the carrier   wave modes. The system’s evolution is then governed through the modulational instability (MI). The latter  signifies the exponential growth of a small plane wave perturbation as it propagates in plasmas. The gain leads to amplification of sidebands, which break up the otherwise uniform wave and lead to energy localization via the formation of localized structures. Thus, the MI   acts as a precursor for the formation of bright envelope solitons, in absence of which we have the formation of dark solitons.
 
Let us now consider the amplitude modulation of a plane drift wave solution of Eq. \eqref{NLSE} of the form $\Phi=\Phi_0e^{-i\Omega_0\tau}$, where $\Omega_0=-Q\Phi_0^2$ with $\Phi_0$ denoting the potential of the drift wave pump.  We then modulate the drift wave amplitude as a plane wave perturbation with frequency $\Omega$ and wave number $K$ as $\Phi=\left(\Phi_0+\Phi_1e^{iK\xi-i\Omega\tau}+\Phi_2 e^{-iK\xi+i\Omega\tau}  \right)e^{-i\Omega_0\tau}$, where $\Phi_{1,2}$ are real constants. Looking for the nonzero solution of the small perturbations, we obtain from Eq. \eqref{NLSE} the following dispersion relation for the modulated drift wave packets:
%16 DR-modulation
\begin{equation}
\Omega^2=(PK^2)^2\left(1-\frac{K_c^2}{K^2}\right),\label{DR-modulation}
\end{equation}
where $K_c= \sqrt{2|Q/P|}|\Phi_0|$ is the critical value of the  wave number of modulation $K$,  such that  MI sets in for $K<K_c$, and the wave will be modulated for $PQ>0$. In the latter, the perturbations grow exponentially during propagation of waves.  On the other hand, for $K>K_c$ the wave is said to be stable $(PQ<0)$ against the modulation. The instability growth  rate is obtained as
%17 instability-rate
\begin{equation}
\Gamma= |P| K^2\sqrt{\frac{K^2_c}{K^2}-1}. \label{instability-rate}
\end{equation}
Clearly, the maximum value of $\Gamma$ is  achieved at $K=K_c/\sqrt{2}$ and is given by $\Gamma_{\text{max}}=|Q||\Phi_0|^2$.  From Eqs. \eqref{P-dispersion} and \eqref{Q-nonlinear} we note that for long wavelength drift modes (with $k<1$), the dispersive coefficient $P$ is always negative, whereas   $Q \gtrless0$ according as
$k\lessgtr1/\sqrt{2}\approx0.71$.  Thus, the drift wave packet is stable  $(PQ <0)$  or unstable $(PQ>0)$  against the  modulation according to when $0< k < 0.71$  or $0.71\lesssim k\lesssim 1$.

Next, we numerically investigate the MI growth rate as shown in Fig. 1. For a fixed $\alpha$ and $\gamma$ (e.g., $\alpha=\gamma=0.05$) we note that as $k$ approaches $1$, the value of $\Gamma$ decreases with a lower cut-off at a lower wave number of modulation.  However, for long wavelength perturbations with $K<1$, the MI growth rate can not be controlled for drift waves with wave numbers $k$ close to $1/\sqrt{2}$. The reason is that when  $k$ approaches $1/\sqrt{2}$, the nonlinear coefficient $Q$ becomes larger and larger.
\begin{figure}
\includegraphics[height=2in,width=3.5in]{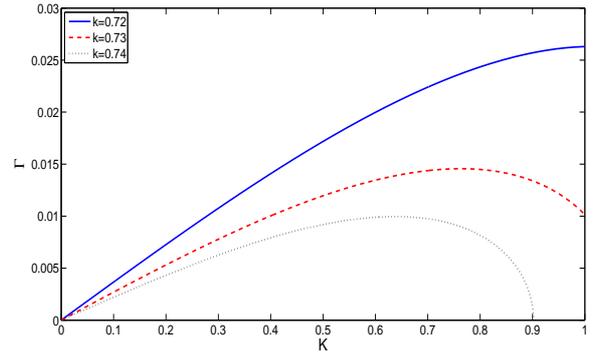}  
\caption{The instability growth rate given by Eq. \eqref{instability-rate} is shown at different wave numbers of the carrier drift mode. The parameter values are $\alpha=\gamma=0.05$.} 
\end{figure}

Excat solutions of the NLSE \eqref{NLSE} can be obtained by considering $\Phi=\sqrt{\Psi}\exp(i\theta)$, where $\Psi$ and $\theta$ are real functions to be determined (see for details, e.g., Refs. \cite{soliton-solutions}). For $PQ>0$ the drift wave is modulationally unstable leading to the formation of   bright envelope modulated wave packets   given by (Fig. 2)
\begin{equation}
\Psi=\Psi_0 \hskip1pt\text{sech}^2\left(\frac{\xi-U\tau}{W}\right),\hskip2pt \theta=\frac{1}{2P}\left[U\xi+\left(\Omega_0-\frac{U^2}{2}\right)\tau\right],\label{bright-envelope}
\end{equation} 
which represents a localized pulse traveling at a speed $U$ and oscillating at a frequency  $\Omega_0$ at rest. The pulse width $W$ is related to the constant amplitude $\Psi_0$ as $W=\sqrt{2P/Q\Psi_0}$.
\begin{figure}
\includegraphics[height=2.5in,width=3.5in]{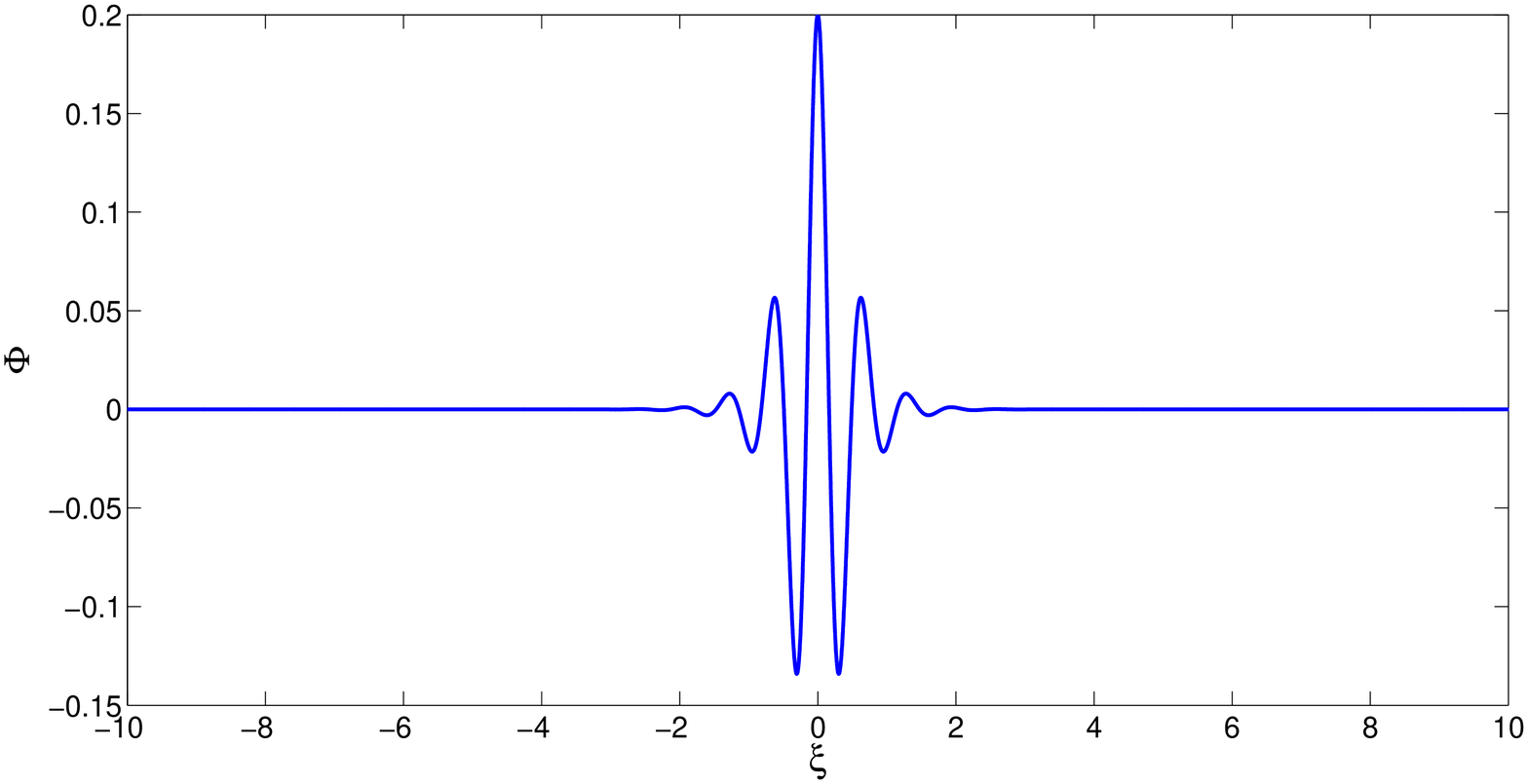}  
\caption{The evolution of the bright envelope soliton as given by Eq. \eqref{bright-envelope}  at $\tau=0$ for $k=0.71$, $\alpha=\gamma=0.05$, $\Psi_0=0.04$ and $U=0.5$.} 
\end{figure}

On the other hand, for $P Q < 0$, the modulationally stable drift wave packet will propagate in the form of a dark envelope soliton   characterized by a depression of the drift wave potential around $\xi =0$. This is given by (Fig. 3)
\begin{eqnarray}
&&\Psi=\Psi_1\hskip1pt \text{tanh}^2\left(\frac{\xi-U\tau}{W_1}\right), \notag \\ &&\theta=\frac{1}{2P}\left[U\xi-\left(\frac{U^2}{2}-2PQ\Psi_1\right)\tau\right],\label{dark-envelope}
\end{eqnarray} 
representing a localized region of hole (void) traveling at a speed $U$. The pulse width $W_2$ depends on the constant amplitude $\Psi_1$ as $W_2=\sqrt{2|P/Q|/\Psi_1}$.
\begin{figure}
\includegraphics[height=2.5in,width=3.5in]{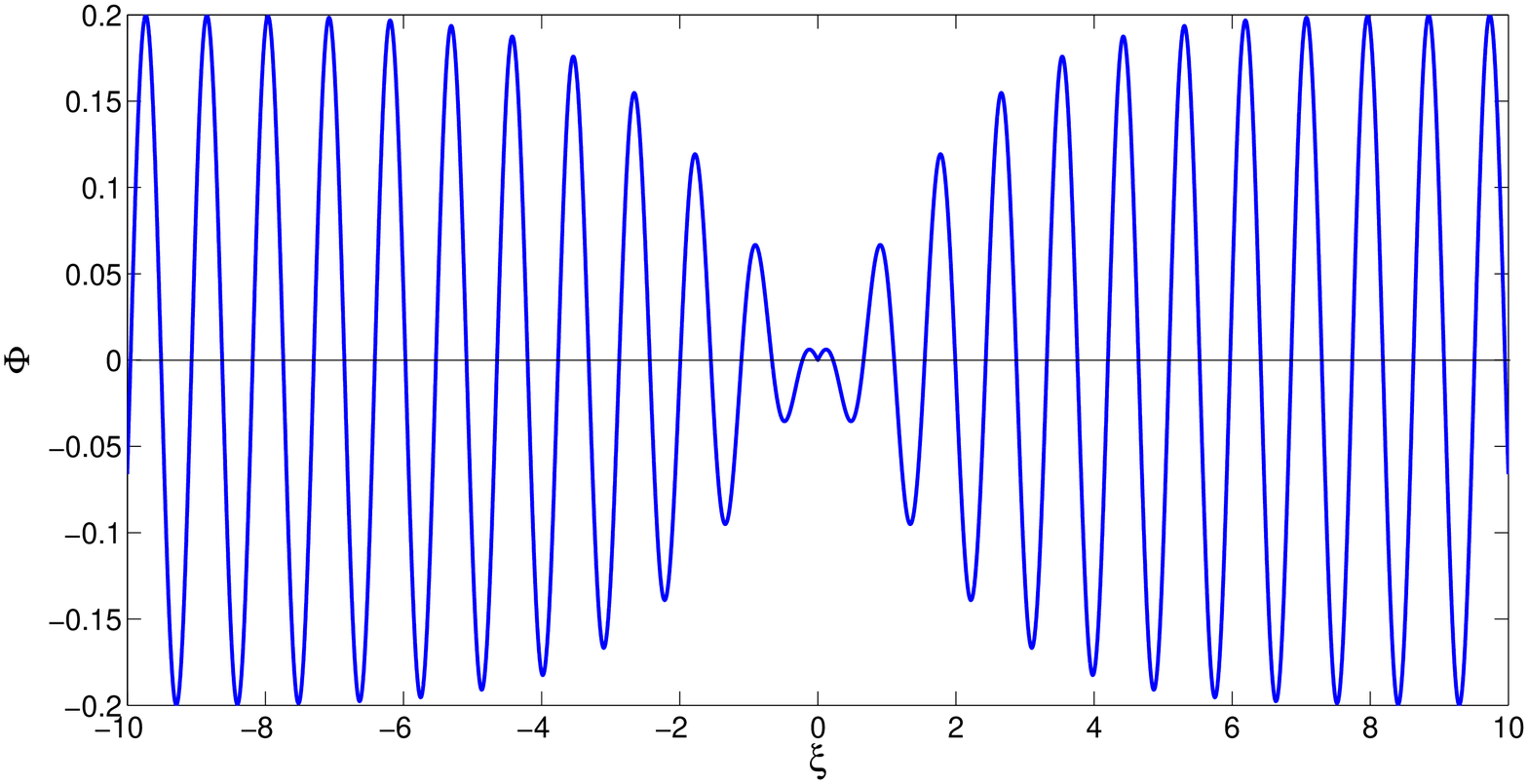}  
\caption{The evolution of the dark envelope soliton as given by Eq. \eqref{dark-envelope}   at $\tau=0$ for $k=0.5$, $\alpha=\gamma=0.05$, $\Psi_0=0.04$ and $U=0.5$.} 
\end{figure}

Furthermore, the NLSE \eqref{NLSE} has a rational solution that is located on a non-zero background and localized both in the $\tau$ and 
$\xi$ directions. For $P Q > 0$ we have the drift rogue waves/rogons \cite{r16,r17}
%18 rogue-solution
%\begin{equation}
%\Phi = \frac{1}{\sqrt{|Q|}}\left[\frac{4(1-2 i \tau)}{1+\tau^2 + 4 \xi^2/|P|} - 1 \right]\exp(i \tau), \label{rogue-solution}
%\end{equation}
\begin{equation}
\Phi = \Phi_0\left[\frac{4\left(1+2 i Q\tau\right)}{1+4Q^2\tau^2 + 2Q \xi^2/P}-1 \right]\exp\left(i Q\tau\right), \label{rogue-solution}
\end{equation}
which reveals that a significant amount of energy is concentrated   in  a relatively small area in  space. The typical form of the rogue wave is shown in Fig. 4.  
Hence, a random perturbation of the drift wave amplitude will grow on account of the modulational instability.
 \begin{figure}
\includegraphics[height=3in,width=3.7in]{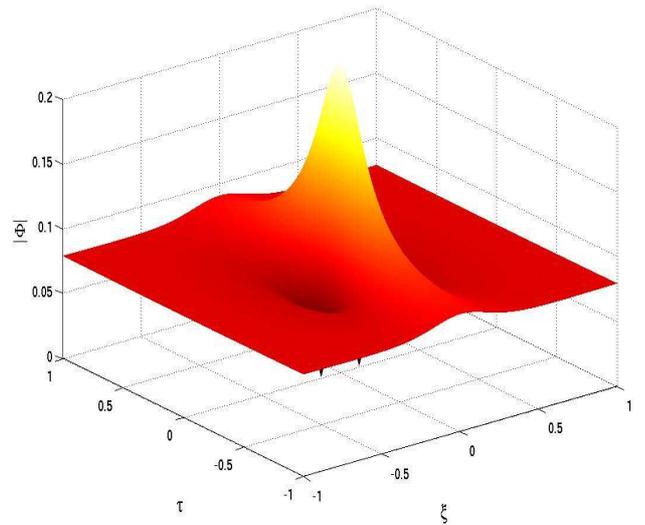}  
\caption{The evolution of the rogon as given by Eq. \eqref{rogue-solution} at $k=0.72$  with parameters $\alpha=\gamma=0.05$.} 
\end{figure} 
%%%%%%
 
To summarize, we have considered the amplitude modulation of a finite amplitude one-dimensonal electrostatic drift wave 
packet in a nonunform magnetoplasma in presence of equilibrium density, electron temperature and magnetic field gradients. It is shown that the dynamics of the modulated drift wave packet is governed by a nonlinear Schr\"odinger 
equation. Since the group dispersion of the drift waves is always negative for $k<1$, the formation of the bright envelope 
drift wave soliton or drift rogons is possible only when $Q < 0$, which happens for $0.71\lesssim k\lesssim 1$. In the opposite case,
viz. $0< k < 0.71$ when $Q > 0$, an amplitude modulated drift wave packet is stable and it propagates in the 
form of a dark envelope soliton \cite{r18}. In conclusion, the present results should be helpful in identifying 
modulated  drift wave packets that may spontaneously emerge in  magnetized space and laboratory plasmas 
that contain equilibrium density, electron temperature and magnetic field inhomogeneities.

\section*{Acknowledgement}
{This research was partially supported by the Deutsche Forschungsgemeinschaft (Bonn) through the project SH 21/3-2 of the Research Unit 1048, and by the SAP-DRS (Phase-II), UGC, New Delhi, through sanction letter No. F.510/4/DRS/2009 (SAP-I) dated 13 Oct., 2009.}

\bibliographystyle{elsarticle-num}

\end{document}